
\font\smaller=cmr8
\magnification=1200
\baselineskip=14truept
\hsize=6truein
\vsize=8.5truein
\nopagenumbers
\parindent=0pt

\def\half{\hbox{$1\over2$}}

\centerline{MECHANICS AS GEOMETRY?}
\vskip 1.5truecm

\centerline{JOHN M. CHARAP}
\centerline{\it Physics Department, Queen Mary and Westfield
College}
\centerline{\it Mile End Road, London E1 4NS, England}

\vskip 5.5truecm
\centerline{ABSTRACT}
\smallskip
\vbox{
\parindent=3truepc
\baselineskip=12truept
\smaller
\narrower
\noindent
\quad
Instead of attempting to give a summary or to identify highlights
of the workshop, the history of the development of analytical
mechanics is outlined, with an emphasis on the themes of \lq
natural motion' and the variational principle. }

\vskip 9truecm
\centerline{Talk given at the Workshop on \lq\lq Constraint Theory and
Quantization Methods"}
\centerline{Montepulciano, Italy, June 1993}
\bigskip
QMW 93-23\hfill August 1993
\vfil
\eject

{\bf 1.^Introduction}
\vskip 0.5truecm

\qquad
Many of the speakers at this Workshop have drawn attention to the
beautiful surroundings in which it was held, and I too would like
to record my appreciation of the choice by the organisers of the
charming theatre in Montepulciano for this second workshop on the
theme of constraints theory and quantisation methods.  All the
participants must be grateful to the sponsors and organisers and
especially to Luca Lusanna for their efforts which have made this
such a stimulating and enjoyable meeting.  When I accepted the
invitation to speak, I had thought that I would be giving the
introductory talk; instead I find myself giving the final talk
which tradition demands should be either a summary, or at least
an overview with particular contributions highlighted.  I propose
to fly in the face of tradition.  The speakers themselves,
expertly controlled by those chairing the sessions, were concise
and to the point.  The abstracts in these proceedings will give
a much better summary than I could offer!  As I listened to the
talks, I put a star against the name of the authors of those
which I thought I might highlight in this talk: the result was
a galaxy of contributions which I thought deserved special
attention, so I will offend none by mentioning none.  What I will
do instead is to revert to my original intention, and simply
outline the historical development of analytical mechanics,
hoping that you too will enjoy being reminded of the antiquity
of the background to our present concerns, and the common cause
we share with our illustrious predecessors in seeking to
understand dynamics as the \lq natural' motion of a system with
constraints.

\qquad
Plato ({\it ca} 429--347 BC) had wished that his pupils should
be first of all well grounded in mathematics, and in particular
geometry, for it is reported that he denied entry to his Academy
to those ignorant of geometry.  For him God geometrises
continually.  The supreme authority for the later Middle Ages in
most branches of learning, certainly in physics, was Aristotle
(384--322 BC).  In his view, natural motion in our (sublunar)
region is when a substance seeks its natural place, and can only
be vertically up or down, whilst for the heavenly bodies natural
motion is in a circle.  He also believed that Nature works in the
shortest possible way, and the straight line is the shortest of
all.  The idea which was to become the Newtonian law of inertia
thus originates with the accepted doctrine of the ancient
philosophers, summarised by Plutarch ({\it ca} 46--127): For
every body is carried according to its natural motion, unless it
be diverted by some other intervening cause.

\qquad
The first of these workshops was held in Arcetri in the hills
above Florence in 1986, and it was in Florence that Galileo
Galilei (1564--1642) wrote in {\it Il Saggiatore} (The Assayer):
\lq\lq Philosophy is written in that great book which ever lies
before our gaze -- I mean the universe -- but we cannot
understand if we do not first learn the language and grasp the
symbols in which it is written.  The book is written in the
mathematical language, and the symbols are triangles, circles and
other geometrical figures, without the help of which it is
impossible to conceive a single word of it, and without which one
wanders in vain through a dark labyrinth."  That mathematics, and
in particular geometry, was needed to understand dynamics was to
him self-evident: it also provided him with a defence -- for a
time -- against the Holy Office.  In 1632, nine years after {\it
Il Saggiatore}, Galileo was compelled to recant his \lq
heresies', and to live out the rest of his life in Arcetri.

\qquad
For Galileo's great contemporary Johannes Kepler (1571--1630)
mathematics was also crucial: \lq\lq For we see that these
motions [of the planets] take place in space and time and this
virtue [force] emanates and diffuses through the spaces of the
universe, which are all mathematical conceptions.  From this it
follows that this virtue is subject also to other mathematical
necessities."   It would be wrong to read into his \lq\lq{\it Ubi
materia, ibi geometria"} a prophecy of General Relativity, but
the linkage between physics and geometry is undeniable.  Likewise
for Ren\'e Descartes (1596--1650): \lq\lq I do not accept or
desire any other principle in Physics than in Geometry or
abstract Mathematics".

\qquad
And so we come to Isaac Newton (1642--1727) whose First Law
\lq\lq Every body continues in its state of rest, or of uniform
motion in a right line, unless it is compelled to change that
state by forces impressed upon it" inherits from the philosophers
of antiquity the notion of natural motion, but makes explicit and
universal what that natural motion is.  The focus of attention
then turns to the forces and their effects on motion.  The
encapsulation of Newton's Laws in the progressively more elegant
and profound formulations of the variational principle underlies
all of classical analytical mechanics, and is embedded in the
foundations of our present interests too.

\qquad
I cannot resist including amongst these quotations from early
times one by Frederick the Great of Prussia (1712--1786): \lq\lq
I have no fault to find with those who teach geometry.  That
science is the only one which does not produce sects; it is
founded on analysis and on synthesis and on the calculus; it does
not occupy itself with probable truth; moreover it has the same
method in every country".

\qquad
In preparing this talk, I have plundered without further
reference three books in particular, listed in the bibliography.
To one of these I am especially indebted, and I use its title for
that of the next section.

\vskip 1truecm
{\bf 2.^The Variational Principles of Mechanics}
\vskip 0.5truecm

\qquad
The configuration of a classical system of $N$ particles may be
represented by a point in a $3N$-dimensional configuration space,
with coordinates $qi$.  The kinetic energy $T$, which can be
written as
$$\eqalignno{
T&=\sum_{\alpha=1}N\half m_\alpha{\bf v}_\alpha2\cr
&={1\over2}g_{ik}{dqi\over dt}{dqk\over dt}&(1).\cr
}$$
suggests a natural way to introduce a Riemannian metric
$$g_{ik}=g_{ik}(q)$$
on configuration space, with line element given by
$$ds2=g_{ik}dqidqk=2Tdt2.\eqno(2)$$
There may be kinematical constraints, which confine the
representative point to a subspace, which may be curved.  Such
constraints will have the form
$$\omega_a:=f_{ai}(q)\,dqi=0,\eqno(3)$$
and are said to be holonomic if the differential form $\omega_a$
is exact.
\smallskip
\qquad
Suppose now that the system is displaced infinitesimally in the
constraint space.  The work done is an invariant differential
form
$$\delta W = F_i\,dqi,\eqno(4)$$
the components of which define the forces (excluding those which
enforce the constraints) acting on the system, as a covariant
vector in the cotangent bundle.  When this form is exact, it is
the differential of what Lanczos calls the work-function
(generalisation for velocity dependent forces is also possible),
and when this is independent of the time $t$ and of the
velocities, it is equated to the negative of the potential energy
$$\delta W = F_i\,dqi= -dV.\eqno(5)$$
Boltzmann (1844--1906) called kinematic conditions involving time
{\it rheonomic}; those which are time-independent are {\it
scleronomic}.  It is only in the latter case that the energy
theorem holds; rheonomic constraints also introduce terms linear
in the generalised veloicities $dqi/dt$ into the expression
Eq.(1) for the kinetic energy and the Riemmanian geometric
picture loses its utility.
\vskip 1truecm
{\it 2.1.^D'Alembert's Principle}
\vskip 0.5truecm
\qquad
In statics, the condition for equilibrium can be embodied in the
principle of virtual work (this was recognised in a primitive
form in Aristotle's derivation of the law of the lever, applied
Stevinus (1598--1620) to pulleys, and by Galileo to the
equilibrium of a body on an inclined plane; and appears in a
recognisably complete formulation in the work of John Bernoulli
(1667--1748)): for any variation of the configuration consistent
with the constraints, the work of the impressed forces is zero.
\qquad
Starting from Newton's Second Law in the form
${\bf F}_\alpha={d\over dt}(m_\alpha{\bf v}_\alpha)$, D'Alembert
(1717--1783) equated the right-hand side of this equation to the
negative of a \lq force of inertia', $-{\bf I}_\alpha$, and so
on writing
$${\bf F}_\alpha + {\bf I}_\alpha =0\eqno(6)$$ allowed the
principle of virtual work to be extended from statics to
dynamics.  When the masses are constant, we have
$$\delta W := \sum_{\alpha}({\bf F}_\alpha-m_\alpha{\bf
a}_\alpha)\cdot\delta{\bf r}_{\alpha} = 0.\eqno(7)$$
\vskip 1truecm
{\it 2.2.^Gauss' Principle of Least Constraint}
\vskip 0.5truecm
\qquad
The equations of motion for the position {\bf r} of a particle
are of second order in time, so the position at time $t+\tau$ is
determined through the equations of motion once the initial
position and velocity, ${\bf r}(t)$ and ${\bf v}(t)$ are
specified.  Using the expansion
$${\bf r}(t+\tau)={\bf r}(t)+{\bf v}(t)\tau+\half{\bf
a}(t)\tau2+\ldots,\eqno(8)$$
suppose that we fix the initial conditions, but vary the
acceleration in an arbitrary way, subject to the constraints.
Then for small $\tau$, we have
$$\delta {\bf r}(t+\tau)=\half\delta{\bf a}(t)\tau2.\eqno(9)$$
Putting this in Eq.(7) gives
$$\sum({\bf F}_\alpha-m_\alpha{\bf
a}_\alpha)\vert_{(t+\tau)}\cdot\delta{\bf a}_\alpha,\eqno(10)$$
or on using the fact that $\tau$ is small, and also noting that
the forces and the masses are not to be varied, we find
$$\delta Z=0,\eqno(11)$$
where what Gauss (1777--1855) called the {\it constraint} is
defined by
$$Z=\sum{({\bf F}_\alpha-m_\alpha{\bf
a}_\alpha)2\over2m_\alpha}.\eqno(12)$$
Note that since the Gauss constraint is non-negative, what he
proposed was a Principle of {\it Least} Constraint.

\vskip 1truecm
{\it 2.3^Hertz's Principle of the Straightest Path}
\vskip 0.5truecm
\qquad
Heinrich Hertz (1857--1894), probably better known today for his
pioneering work on electromagnetic oscillations, also made
interesting contributions to theoretical mechanics.  He tried to
develop a force-free mechanics, of which more below.  But he
earlier recognised that in the case when the forces (other than
those which impose the constraints) vanish, the Gaussian
constraint may be interpretted as the geodesic curvature of the
trajectory of the system in configuration space, with the
geometry given by the line element of Eq.(2).  So the trajectory
is the straightest path in the subspace determined by the
constraints.

\vskip 1truecm
{\it 2.4^The Action Principle}
\vskip 0.5truecm
\qquad
One may think of D'Alembert's Principle as a variational
principle.  But its application is rendered difficult by the fact
that in general $\delta W$ is not an exact differential.  It was
Maupertius (1698--1759) who introduced the notion of the {\it
action} as a quantity which was minimised in any dynamical
process, but his arguement was flawed, and had indeed been
anticipated -- correctly -- by Euler (1707--1783).  In the work
of Lagrange (1736--1813) the action principle, including the
method of undetermined multipliers for handling constraints,
first appears in its general form.  By integrating the variation
$\delta W$ of the D'Alembert principle with respect to time one
obtains
$$\eqalignno{\int_{t_i}{t_f}\,\sum_\alpha{\bf
F}_\alpha\cdot\delta{\bf r}_\alpha&=-\int_{t_i}{t_f}\,\delta
V\,dt&(13)\cr
\noalign{\hbox{and}}\cr
-\int_{t_i}{t_f}\,\sum_\alpha{d{\bf p}_\alpha\over
dt}\cdot\delta{\bf r}_\alpha&=-\Bigl[\sum_\alpha{\bf
p}_\alpha\cdot
\delta{\bf r}_\alpha\Bigr]_{t_i}{t_f}
+\int_{t_i}{t_f}\,\sum_\alpha{\bf p}_\alpha
\cdot\delta{\bf v}_\alpha\,dt\cr
&=-\Bigl[\sum_\alpha{\bf p}_\alpha\cdot
\delta{\bf r}_\alpha\Bigr]_{t_i}{t_f}+\int_{t_i}{t_f}\,\delta
T\,dt,&(14)\cr
}$$
so that with $L=T-V$,
$$\int_{t_i}{t_f}\,\delta W\,dt=\int_{t_i}{t_f}\,\delta L\,dt
-\Bigl[\sum_\alpha{\bf p}_\alpha\cdot
\delta{\bf r}_\alpha\Bigr]_{t_i}{t_f}.\eqno(15)$$
The result is then
$$\delta A=0,\eqno(16)$$
where the action $A$ is
$$A=\int L\,dt,\eqno(17)$$
and the variations are such that the initial and final
configurations are unaffected.

\qquad
This variational principle famously results in Lagrange's
equations:
$${d\over dt}{\partial L\over\partial\dot qi}={\partial
L\over\partial qi}.\eqno(18)$$

\vskip 1truecm
{\it 2.5^Hertz's Proposal to Eliminate Force}
\vskip 0.5truecm

\qquad
Hertz was critical of the action principle, first because of its
restriction to holonomic systems, and secondly because of the
logical difficulties he saw in the concept of force.  He wished
to base mechanics on only three independent fundamental
conceptions; space, time and mass.  His idea was to assume that
there were \lq concealed masses' described by what J.J. Thomson
referred to as \lq kinosthenic' variables -- a variable $qa$
(say) is said to be kinosthenic when only its velocity $\dot
qa$, and not the coordinate $qa$ itself, enters into the
Lagrange function $L$.  Then since
$${\partial L\over\partial qa}=0,$$
the conjugate momentum
$$p_i:={\partial L\over\partial\dot qi}\qquad (i=a)\eqno(19)$$
is constant in time.  The equation $p_a=c_a$ is then to be solved
so as to express $\dot qa$ in terms of $c_a$ and the other
variables.  On writing
$$\bar L:=L-c_a\dot qa\eqno(20)$$
all reference to $qa$ as well as to $\dot qa$ has disappeared.

\qquad
However, the non-kinosthenic velocities $\dot qi\quad,i\neq a$
enter into the modified Lagrangian $\bar L$ differently from in
the original $L$; in general there will be terms {\it linear} in
the velocities, even if there were no such (so-called {\it
gyroscopic}) terms in $L$.  And there will also be a modification
to the velocity-independent part of the Lagrangian, that is to
say, to the potential.  What Hertz conjectured was that {\it all}
of the potential energy might arise in this way, from elimination
of kinosthenic variables describing the motion of his concealed
masses.  I do not myself believe that this idea is sustainable,
but it does have a certain resonance with what is done in many
situations where an effective interaction is obtained by
integrating out the dependence on some unobserved variables.

\vskip 1truecm
{\it 2.6^Time as a Kinosthenic Variable: Jacobi's Principle}
\vskip0.5cm
\qquad
In the familiar expression for the action
$$A=\int L(qi,\dot qi)\,dt\eqno(21)$$
it is possible to regard the time $t$ as an extra coordinate, by
choosing some other parameter, $\tau$ say, for the path
connecting the initial and final configurations, and then with
a prime denoting differentiation with respec to $\tau$, we obtain
$$A=\int\,\tilde L\,d\tau,\eqno(22)$$
where
$$\tilde L=L(qi,q{i\prime}/t\prime)t\prime.\eqno(23)$$
We may suppose that the original Lagrangian $L$ did not depend
explicitly on $t$, it was scleronomic.  Then in $\tilde L$, the
variable $t$ enters only through its derivative $t\prime$, and
so is kinosthenic.  Its conjugate momentum is then a constant,
$$p_t=L-p_i\dot qi=\hbox{constant}=-E.\eqno(24)$$
We are to form the modified Lagrangian $\bar L$ as above,
$$\bar L:=\tilde L-p_tt\prime=p_i\dot qit\prime,\eqno(25)$$
and then eliminate $t\prime$ from the modified action
$$\bar A=\int\bar L\,d\tau\eqno(26)$$
using Eq.(24).  When the kinetic energy has the form Eq.(1), this
leads to the expression
$$\bar A=\int\sqrt{2(E-V)}\,ds\eqno(27)$$
for the action, as introduced by Jacobi (1804--1851).  The
variational principle $\delta\bar A=0$ now gives the {\it
trajectory}, but does not use time as the parameter along the
trajectory (although of course time may be re-introduced if one
so wishes).

\qquad
The Jacobi form of the action, $\bar A$, can also be thought of
as the length of the path connecting the initial to the final
configuration in $q$-space, with a metric now different from that
introduced in Eq.(1).  Instead we have
$$d\sigma:=\sqrt{2(E-V)}\,ds\eqno(28)$$
and
$$\bar A=\int\,d\sigma.\eqno(29)$$

\vskip 1truecm
{\it 2.7^The Canonical Form for the Action}
\vskip 0.5truecm

\qquad
 A transformation of the kind introduced by Legendre (1752--1833)
in his study of systems of partial differential equations allowed
Hamilton (1805--1865) to go from the Lagrange expression for the
action $A=\int L\,dt$ to the new form
$$A=\int\Bigl[\sum_i\,p_i\dot qi - H(q,p)\Bigr]\,dt,\eqno(30)$$
in which the focus of attention has been redirected from the
configuration space of the $q$-variables to the phase space [the
name is due to Gibbs (1839--1903)] of the $q$- and $p$-variables
together.  The principle of stationary action now leads to a
motion of the phase points $(qi, p_i)$ as the \lq Hamiltonian
flow'.  By introducing an additional coordinate -- time -- to the
$2n$-dimensional phase space, Elie Cartan (1869--1951) pictured
the motions of the dynamical system as a manifold of
non-intersecting curves in what he called \lq state space'.  But
we can go further, and again regard time as an additional
coordinate, say $t=q0$, and introduce also a conjugate momentum
$p_0$, so that the action integral is now
$$A=\int\,\sum_{i=0}n\,p_iq{i\prime}\,d\tau.\eqno(31)$$
What has happened to the dynamics?  For in this expression for
the action the Hamiltonian function does not appear.  The answer
is that the variation of this expression for the action is
subject to a constraint, the Hamiltonian constraint
$$K:=p_0+H(q,p)=0.\eqno(32)$$

\qquad
Variation of the canonical expression Eq.(31) for the action with
the constraint Eq.(32) imposed through the Lagrange $\lambda$-
method  means making stationary the integral
$$\bar A=\int_{\tau_i}{\tau_f}\,\Bigl(\sum_{i=0}n
p_iq{i\prime}-\lambda K\Bigr)\,d\tau,\eqno(33)$$
and leads directly to Hamilton's equations if the identification
of $t, \tau$ and $q0$ is made.

\qquad
If instead the kinosthenic variable $t=q0$ is eliminated, the
result is to give the Lagrange formulation.  The procedure is to
write $p_0$ as a constant ($=-E$) in equation Eq.(32), and then
rewrite it as
$${T\prime\over t{\prime2}}+V=E,\eqno(34)$$
where $T\prime=\half g_{ij}q{i\prime}q{j\prime}=Tt{\prime2}$,
which is to be used to eliminate $t\prime$ from the modified
form
$$\bar A=\int\Bigl(\sum_{i=1}n p_iq{i\prime}\Bigr)\,d\tau$$
of the action.  This may be done by imposing Eq.(34) as a
constraint, leading to
$$\tilde A=\int\Bigl(\sum_{i=1}n
p_iq{i\prime}+\lambda\bigl({T\prime\over
t{\prime2}}+V\bigr)\Bigr)\,d\tau.\eqno(35)$$
Varying with respect to $t\prime$ gives $\lambda=-t\prime$,
since the sum involves $t\prime$ through
$\sum=2T\prime/t\prime$ (plus another term independent of
$t\prime$ in the case when gyroscopic terms are present, which
has no effect on the conclusion), and this then results in
$$\tilde
A=\int_{\tau_i}{\tau_f}Lt\prime\,d\tau=\int_{t_i}{t_f}L\,dt
.\eqno(36)$$

\qquad
Alternatively, imposing the constraint through
$${\half g{ij}p_ip_j\over E-V}=1\eqno(37)$$
leads to Jacobi's formulation (we have omitted gyroscopic terms
for the sake of simplicity).  If this condition Eq.(37) is
imposed through the $\lambda$-method, and the result re-expressed
in configuration space, it leads to the action integral
$$A=\int_{\tau_i}{\tau_f}(E-
V)g_{ij}q{i\prime}q{j\prime}\,d\tau\eqno(38)$$
which resembles the Jacobi form, but without the square-root.
[You may have seen something similar done in relating the
Polyakov and the Nambu-Goto forms of the string action.]

\vskip 1truecm
{\it 2.8^Hamilton-Jacobi Theory}
\vskip 0.5truecm

\qquad
The time evolution of a mechanical system may be regarded as the
continuous unfolding of a canonical transformation generated by
Hamilton's Principle Function.  This function is nothing other
than the action integral along the trajectory in configuration
space; it can also be regarded as the length of the trajectory
as determined by the line-element
$d\sigma=Ldt.$

\qquad
The generating function is
$$\eqalignno{
W_{if}&=W(q1_i,\ldots,qn_i,t_i;q1_f,\ldots,qn_f,t_f)\cr
&=\int_{t_i}{t_f}\,L\,dt\cr
&=\int_{\tau_i}{\tau_f}\,L\,t\prime\,d\tau.&(39)\cr}$$
And it is equal to the stationary value of the canonical integral
$$A=\int_{t_i}{t_f}\,\sum_{j=1}n p_j\dot qj\,dt\eqno(40)$$
when variations are restricted to the energy-surface
$$H(q1,\ldots,qn;p_1,\ldots,p_n)-E=0;\eqno(41)$$
or alternatively to
$$A=\int_{\tau_i}{\tau_f}\,\sum_{j=0}n p_j
q{j\prime}\,d\tau\eqno(42)$$
with the constraint
$$H(q1,\ldots,qn;p_1,\ldots,p_n)+p_0=0.\eqno(43)$$
All of dynamics is here encapsulated in the constraint equation.

\vskip 1truecm
{\bf 3.^Quantization Methods}
\vskip 0.5truecm

\qquad
This workshop has of course been concerned with the problem of
constraints over and beyond the Hamiltonian constraint.  It has
also been concerned with quantization methods, and I want to add
a few words on this, in the form of a personal reminiscence.  I
had the good fortune to attend the Symposium on \lq\lq The
Physicist's Coception of Nature", held in 1972 at the ICTP,
Trieste to celebrate the seventieth birthday of P.A.M. Dirac
(1902--1984).  One of the contributions was a paper by B.L. van
der Waerden \lq From Matrix Mechanics and Wave Mechanics to
Unified Quantum Mechanics', in which he discussed a letter
written in 1926 by Wolfgang Pauli to Pascual Jordan.  The letter
made passing (dismissive) reference to work by Cornelius Lanczos,
also referred to in the 1926 paper in which Erwin Schr\"odinger
demonstrated the equivalence between matrix and wave mechanics.
Lanczos had in 1925 (!) submitted a paper to {\it Zeitschrift
f\"ur Physik} in which he proposed an {\it integral} equation,
the eigen-values of which should be the reciprocal energy values
-- this would in fact be the integral-equation form of
Schr\"odinger's equation $H\psi=E\psi$.  Van der Waerden showed
that Lanczos' paper deserved to be better recognised.  What he
did not appreciate was that Lanczos himself deserved to be better
recognised.  For when Leon Rosenfeld (whose knowledge of the
contemporary world of physics deriving from his editorship of
{\it Physics Letters} was probably unparalleled) as chairman of
the session announced that \lq ... Cornelius Lanczos is here and
has heard this very well deserved vindication of his contribution
at that time', van der Waerden, in common with most of his
audience, was quite taken aback; he exclaimed \lq Oh, you are
Lanczos?'  It was, as the symposium proceedings say, an historic
encounter.

\qquad
I have told this story because it helps me to repay my debt to
Lanczos for the plunder I have taken from his book.  It also
allows me to turn my attention to Dirac, whose work on
constraints was of such seminal importance.  Less well known
perhaps is the position of Dirac's work in the formulation of
quantum mechanics by Julian Schwinger and especially by Richard
Feynman.  Both of them base their approach on what amounts to a
quantum form of the action principle (in Schwinger's case, based
on the Hamiltonian form Eq.(30), whilst Feynman uses the
Lagrangian form Eq.(21)).  And this is in fact to be found in a
brief but complete formulation in Dirac's famous text-book {\it
Quantum Mechanics} -- it is \S 32, which Dirac in a footnote
suggests \lq may be omitted by the student who is not specially
concerned with higher dynamics'.  But not by us!

\vskip 1truecm
{\bf 4.^Bibliography}
\vskip 0.5truecm

\vbox{\parindent=0.8truecm\item{}
Cornelius Lanczos, {\it The Variational Principles of Mechanics},
(University of Toronto Press, Toronto, 1949).

\item{}Max Jammer, {\it Concepts of Force: a study in the
foundations of dynamics}, (Harvard University Press, Cambridge,
Mass., 1957).

\item{}Alan L. Mackay, {\it The Harvest of a Quiet Eye: a
selection of scientific quotations}, (The Institute of Physics,
Bristol and London, 1977).}

\bye